\documentclass[aps, amssymb, showpacs, superscriptaddress, twocolumn]{revtex4-1}

\usepackage[version=3]{mhchem} 
\usepackage[T1]{fontenc} 
\usepackage{graphicx}
\usepackage{amssymb, gensymb}
\usepackage{xfrac}
\usepackage{xcolor}

\begin{document}

\title{Enhanced Thermoelectric Properties in a New Silicon Crystal Si$_{24}$ with Intrinsic Nanoscale Porous Structure}

\author{Kisung Chae}
\thanks{Contributed equally to this work}
\affiliation{Korea Institute for Advanced Study, Seoul, South Korea}

\author{Seoung-Hun Kang}
\thanks{Contributed equally to this work}
\affiliation{Korea Institute for Advanced Study, Seoul, South Korea}

\author{Seon-Myeong Choi}
\email{Present address: Samsung Advanced Institute of Technology, Suwon, South Korea}
\affiliation{Korea Institute for Advanced Study, Seoul, South Korea}

\author{Duck Young Kim}
\email{duckyoung.kim@hpstar.ac.cn}
\affiliation{Center for High Pressure Science \& Technology Advanced Research, Shanghai, P. R. China}
\affiliation{Division of Advanced Nuclear Engineering, POSTECH, Pohang, South Korea} 

\author{Young-Woo Son}
\email{hand@kias.re.kr}
\affiliation{Korea Institute for Advanced Study, Seoul, South Korea}


\date{\today}

\begin{abstract}
Thermoelectric device is a promising next-generation energy solution owing to its capability to transform waste heat into useful electric energy, which can be realized in materials with high electric conductivities and low thermal conductivities. A recently synthesized silicon allotrope of Si$_{24}$ features highly anisotropic crystal structure with nanometre-sized regular pores. Here, based on first-principles study without any empirical parameter, we show that the slightly doped Si$_{24}$ can provide an order-of-magnitude enhanced thermoelectric figure of merit at room temperature, compared with the cubic diamond phase of silicon. We ascribe the enhancement to the intrinsic nanostructure formed by the nanopore array, which effectively hinders heat conduction while electric conductivity is maintained. This can be a viable option to enhance the thermoelectric figure of merit without further forming an extrinsic nanostructure. In addition, we propose a practical strategy to further diminish the thermal conductivity without affecting electric conductivity by confining rattling guest atoms in the pores.
\end{abstract}

\maketitle

\section{Introduction}
Thermoelectricity is one of the renewable energy solutions that converts waste heat into viable electric energy without generating greenhouse gases. It creates electric current by thermal motion of charge carriers due to a temperature gradient. Performance of thermoelectric materials can be evaluated through dimensionless figure of merit (ZT), 
\begin{equation}
ZT = \dfrac{\sigma S^2}{\kappa_{l}+\kappa_{e}}T,
\label{eq:1}
\end{equation}
which is determined by Seebeck coefficient ($S$), electric conductivity ($\sigma$) and thermal conductivities ($\kappa$) at a given temperature ($T$). For $\kappa$, subscripts $l$ and $e$ refer to lattice and electronic contributions to heat conduction, respectively. To increase the ZT value, a material needs to be a good electrical conductor as well as a good heat insulator as seen in Eq.~(\ref{eq:1}). Such a conflicting condition has recently been realized in a layered crystalline material~\cite{Zhao:2014du, Zhao:2016hs} with strong bond anharmonicity, which can be measured by Gr\"uneisen parameter and be a fingerprint for potential thermoelectric materials.

However, it is still challenging to achieve a high ZT in three-dimensional (3D) crystals because the materials properties listed above are coupled to each other. For instance, increasing a charge carrier in 3D semiconductor by doping would increase both $\sigma$ and $\kappa_{e}$ simultaneously. Also, lifetimes of both charge carriers and phonons in a heavily doped material will be significantly lowered due to increased scattering. In order to overcome this complication, various efforts have been made to reduce the $\kappa_{l}$ by post-processings such as alloying~\cite{Dughaish:2002ei, Yamashita:2002fp}, nanostructuring~\cite{Lee:2008bi, He:2011fh, Boukai:2008iz, Hochbaum:2008hl, Lee:2016im, Joshi:2008dd, Wang:2008te, Tang:2010kv, Nakamura:2015jp}, and confining rattling guest atoms in cages of skutterudites and clathrates~\cite{Dolyniuk:2016ht}. The alloy and nanostructure, however, might lower the mobility of charge carriers by enhancing impurity/boundary scattering, and the confinement is only valid for porous materials which can host guest atoms.

Aforementioned favorable conditions for a higher ZT can be found in a new silicon allotrope, Si$_{24}$, with intriguing electronic properties such as quasi-direct bandgap~\cite{Kim:2014el}. Specifically, Si$_{24}$ shows the \emph{intrinsic} nanostructure as in the case of a good thermoelectric material, SnSe~\cite{Zhao:2014du, Zhao:2016hs} [see Figure~\ref{fig1}(a) and (b)]. This is in contrast to the previous studies for Si with post-processings of \emph{extrinsic} nanoscale structures~\cite{Lee:2008bi, He:2011fh, Boukai:2008iz, Hochbaum:2008hl, Lee:2016im, Joshi:2008dd, Wang:2008te, Tang:2010kv, Nakamura:2015jp}. Moreover, the Si$_{24}$ features regular array of nanosized pores as in skutterudites and clathrates, which makes it even more promising as a thermoelectric material. This indicates that an enhanced thermoelectric property can be achieved in a bulk Si material without post-processings, rendering thermal stability in operating conditions. We note here that Si$_{24}$ is shown to be stable in a wide range of temperature~\cite{Kim:2014el} ($\sim$750 K) and pressure~\cite{PhysRevB.95.094306} ($\sim$8 GPa). We also point out some of the advantages of employing a Si material over other materials containing heavy metal elements (e.g., PbTe) in that Si is cheap, non-toxic, Earth-abundant and free of phase separation.

Besides, due to the complex interactions involved in the transport phenomena as mentioned above, describing transport properties to a reasonable degree remains challenging. Some of the previous studies~\cite{Zhang:2016bh, Ouyang:2017bs} used too simplified approximations to the electronic relaxation times. For instance, overestimated values of ZT were reported for black phosphorous obtained with a single-valued relaxation time, adapted from the experiment~\cite{Zhang:2016bh}, from a deformation potential theory~\cite{Qin:2014kc} or from a constant relaxation time approximation~\cite{Zhang:2014if}. When the effects of electron-phonon interactions were explicitly taken into account, the ZT values decreased by orders of magnitude~\cite{PhysRevB.91.235419}. 

In this Letter, we report the enhanced thermoelectric properties of the Si$_{24}$ over the cubic diamond phase counterpart (dSi) by performing various first-principles calculations without using empirical relaxation times of electron and phonon in Si$_{24}$. Specifically, all of the elastic scattering events between electron and phonon with varying energies and momenta were explicitly enumerated for electric conductivity, $\sigma$. Similarly, the anharmonic effects of heat transport due to the three-phonon scattering were explicitly considered for the lattice thermal conductivity, $\kappa_{l}$. We ascribe the order-of-magnitude enhancement in the ZT to a significant anharmonicity in Si$_{24}$, which is also confirmed by anomalously high Gr\"{u}neisen parameters. In synergy with rattling effects of guest atoms in the nanopores, we believe that the Si$_{24}$ can serve as a promising thermoelectric material.

\section{Computational Details}
To describe the reliable thermoelectric properties for Si$_{24}$ using first-principles calculations, we performed density functional theory calculations as implemented in Quantum Espresso package~\cite{Giannozzi:2009vu}. We used norm-conserving pseudopotentials~\cite{Troullier:1991ey} with the plane wave kinetic energy cutoff of 60 Ry (816.34 eV), and exchange-correlation functional of Perdew-Burke-Ernzerhof~\cite{Perdew:1996iq} (PBE) within generalized gradient approximation was used. In order to evaluate the $\sigma$ explicitly from first-principles calculations, the phonon-mediated $\sigma$ was evaluated by using the transport Eliashberg spectral function~\cite{Allen:1978kg, Park:2014ht} as implemented in EPW package~\cite{Noffsinger:2010hw}. The Bloch wave function calculated on a $\Gamma$-centered $\boldsymbol{k}$/$\boldsymbol{q}$-point mesh of 10$\times$10$\times$6 was interpolated to a much denser grid of 40$\times$40$\times$9 by using the maximally localized Wannier functions (MLWFs)~\cite{Marzari:2012eu, Souza:2001ba, Marzari:1997co}. Using the MLWFs, the relaxation time ($\tau^{e-ph}_{n,\boldsymbol{k}}$) by electron-phonon coupling were computed on the denser $\boldsymbol{k}$/$\boldsymbol{q}$ grids as
\begin{widetext}
\begin{align}
\label{eq:2}
\nonumber \frac{1}{\tau^{e-ph}_{n,\boldsymbol{k}}} = \frac{2\pi}{N}\sum_{m,\lambda,\boldsymbol{q}}|g^{\boldsymbol{k},\lambda}_{m,n}|^{2} \left[ \delta\left(\varepsilon_{m,\boldsymbol{k}}-\varepsilon_{n,\boldsymbol{k}}+\omega_{\lambda,\boldsymbol{q}}\right) \left(n_{\lambda,\boldsymbol{q}}+f_{m,\boldsymbol{k}}\right) \right. + \\
\left. \delta\left(\varepsilon_{m,\boldsymbol{k}}-\varepsilon_{n,\boldsymbol{k}}-\omega_{\lambda,\boldsymbol{q}}\right) \left(n_{\lambda,\boldsymbol{q}}-f_{m,\boldsymbol{k}}+1\right) \right],
\end{align}
\end{widetext}
where $\omega_{\lambda,\boldsymbol{q}}$ is the energy of a phonon with polarization $\lambda$ and wave vector $\boldsymbol{q}$ in the Brillouin zone (BZ), and $n_{\lambda,\boldsymbol{q}}$ and $f_{m,\boldsymbol{k}}$ are Bose-Einstein and Fermi-Dirac distribution functions, respectively. $N$ is the total number of $\boldsymbol{k}$/$\boldsymbol{q}$ grid points in the full BZ. The $\varepsilon_{n,\boldsymbol{k}}$ is energy of a Bloch state at $n$th band at $\boldsymbol{k}$ point in the BZ~\cite{PhysRevB.76.165108}. The $g^{\boldsymbol{k},\lambda}_{m,n}$ is an electron-phonon coupling matrix element, which is computed as $\langle {\Psi_{m,\boldsymbol{k}+\boldsymbol{q}}}|\partial_{\lambda,\boldsymbol{q}}V|{\Psi_{n,\boldsymbol{k}}}\rangle$. $\Psi_{n,\boldsymbol{k}}$ is a Kohn-Sham wave function, and $\partial_{\lambda,\boldsymbol{q}}V$ is the variation of the Khon-sham potential for a unit displacement of the nuclei along the phonon mode of polarization $\lambda$ and wave vector $\boldsymbol{q}$. Using the $\tau^{e}_{n,\boldsymbol{k}}\approx\tau^{e-ph}_{n,\boldsymbol{k}}$ as in Eq.~(\ref{eq:2}), electron transport coefficients were calculated by using bolztrap code~\cite{Madsen:2006fj} within semi-classical Boltzmann transport equation: i.e., $\sigma_{\alpha\beta}\left(\varepsilon\right)=\frac{e^2}{N}\sum\limits_{n,\boldsymbol{k}}\tau^{e-ph}_{n,\boldsymbol{k}}v_{\alpha,n,\boldsymbol{k}}v_{\beta,n,\boldsymbol{k}}\left(-\frac{\partial f_{n,\boldsymbol{k}}}{\partial\varepsilon}\right)$, where $\alpha$, $\beta$=$x$, $y$ or $z$, and $N$ is the number of $\boldsymbol{k}$-points in the full Brillouin zone. The $v_{\alpha,n,\boldsymbol{k}}$ is a group velocity of electrons at $\varepsilon_{n,\boldsymbol{k}}$ along $\alpha$. 
Note that other contributions for $\sigma$ such as electron-electron and impurity scattering are assumed to be negligible (see SI S1).
The Seebeck coefficient, S, shown in Eq.~(\ref{eq:1}) was calculated from the Mott formula at the Fermi level as 
$$S = \dfrac{\pi^2 k^2_B T}{3e} \left. \dfrac{\mathrm{d}\left(\log\left(\sigma\right)\right)}{\mathrm{d}\varepsilon}\right|_{\varepsilon=\varepsilon_F},$$ 
where $e$ and $k_B$ are the charge of an electron and Boltzmann's constant, respectively. The electronic contribution to the thermal conductivity ($\kappa_{e}$) was evaluated \emph{via} Wiedemann-Franz law~\cite{Ziman:1962vn}: $\kappa_{e} = L_{0}\sigma T$ with $L_{0}$=2.44$\times$10$^{-8}$ W$\cdot$ohm/K$^2$.

Similarly, the lattice thermal conductivity ($\kappa_{l}$) with anharmonic effects between three phonons was computed by solving Boltzmann transport equation as implemented in phono3py code~\cite{Togo:2015dy}. The third-order interatomic force constants were evaluated for dynamical matrix by a finite displacement method with a displacement and distance cutoff of 0.03~\AA\, and 15~\AA\, respectively. In a (3$\times$3$\times$2) supercell with 216 Si atoms, a total of 12,108 displaced configurations were generated. Hellmann-Feynman forces in each configuration was computed by first-principles calculations by using Vienna \emph{ab initio} software package~\cite{Kresse:1993bz, Kresse:1994il, Kresse:1996kg, Kresse:1996kl} with projector augmented wave method~\cite{Kresse:1999dk} and PBE exchange-correlation functional~\cite{Perdew:1996iq}. Plane wave was expanded with the kinetic cutoff energy of 400 eV, and $\Gamma$-centered 9$\times$9$\times$3 grid points in the BZ were used for the primitive cell of Si$_{24}$. After all the force evaluation, the $\kappa_{l}$ was evaluated by integrating over $\boldsymbol{q}$ points in the full BZ of 14$\times$14$\times$14 grid points.

\section{Results and Discussion}
\begin{figure}[]
	\centering
	\includegraphics[width=\columnwidth]{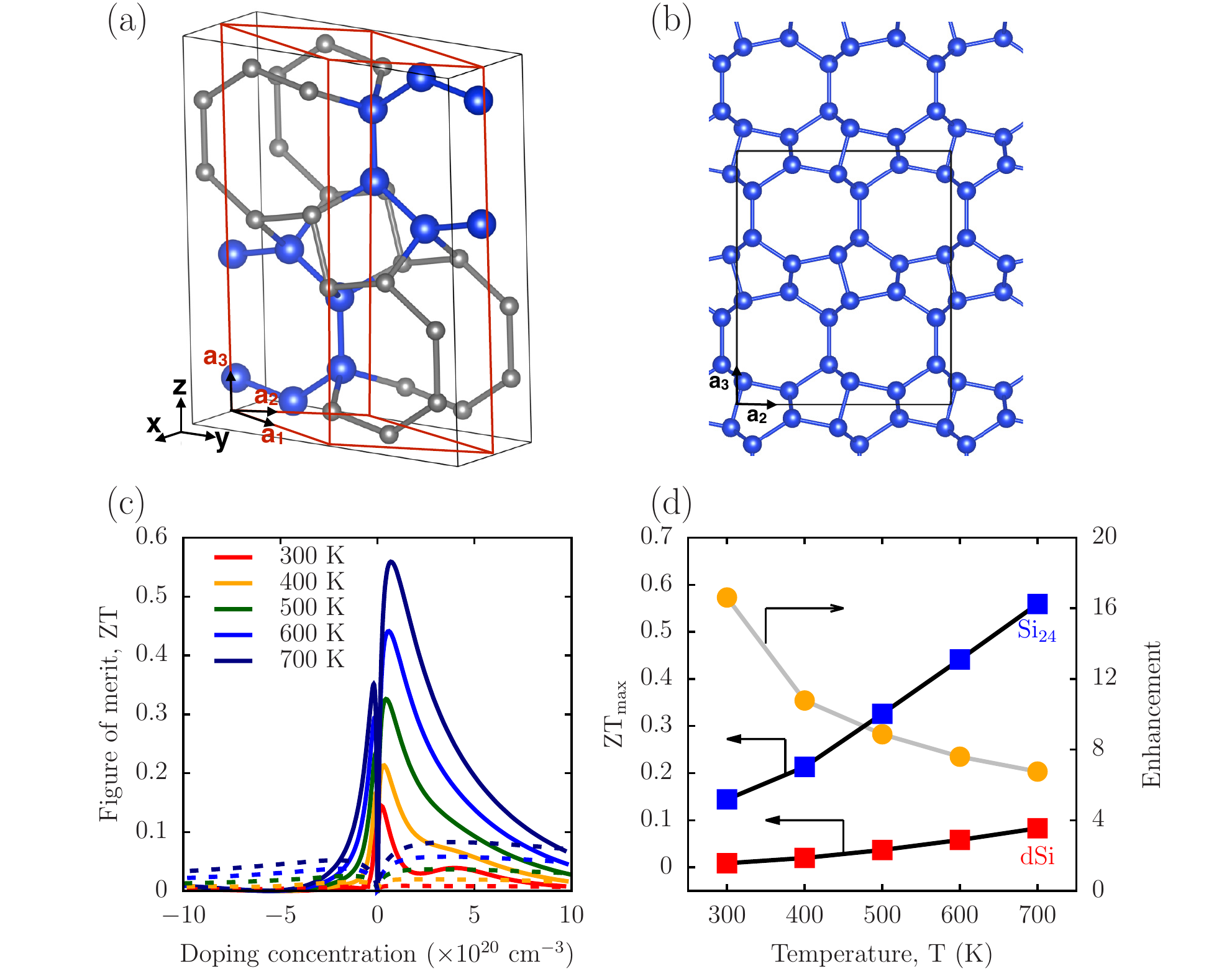}
	\caption{Ball-and-stick models of the Si$_{24}$ crystal with (a) a perspective view and (b) orthogonal projection along the $x$-axis. (a) The primitive cell and the orthorhombic Bravais unitcell are shown as the red and black parallelepipeds, respectively. Twelve atoms in the primitive cell are shown in large blue balls, while the rest of the atoms are shown as small gray balls. (b) Continuous nanopores are shown where a one-dimensional array of Na ions (not shown) were confined for synthesis. (c) Thermoelectric figure of merit (ZT) of Si$_{24}$ along $x$-axis are shown in continuous lines as a function of temperature, and the ZT values for dSi are also shown as dashed lines for comparison. Positive and negative doping concentrations ($n$) refer to that of excess electron and hole, respectively. (d) The maximum values of ZT (ZT$_{\mathrm{max}}$) for a varying temperature are shown as squares, of which the optimum doping concentrations (in 10$^{20}$ cm$^{-3}$) for Si$_{24}$ are 0.21, 0.35, 0.44, 0.59 and 0.69 for increasing temperature; the values are 0.97, 1.81, 3.12, 3.43 and 4.03 for dSi. The enhancement (ratio of the ZT$_{\mathrm{max}}$ for Si$_{24}$ to dSi) is also plotted as orange circles.}
	\label{fig1}
\end{figure}

The Si$_{24}$ crystal features an open-channel structure along a crystallographic axis ($x$) as shown in Figure~\ref{fig1}(a) and (b). Each of the channels, of which the diameter is approximately 0.54 nm, is composed of eight-membered-rings of distorted sp$^{3}$ Si bonds. When compared to the dSi having the ideal sp$^3$ bonding character with the distance of 2.35 \AA\, and with the angle of 109.5$\degree$, those in Si$_{24}$ range from 2.33 \AA\, to 2.41 \AA, and from 97.73$\degree$ to 135.63$\degree$, respectively~\cite{Kim:2014el}. In particular, the open-channel structure is attributed to its two-step synthesis process such that a high pressure phase of Na$_{4}$Si$_{24}$ compound is initially synthesized and pressurized at a hydrostatic pressure of 10 GPa, and that only loosely captured Na atoms in the cage are selectively removed from the compound at ambient pressure~\cite{Kim:2014el}. As a result, the structure with a regular array of one-dimensional nanopores is retained as seen in Figure~\ref{fig1}(b). Note that the structure is highly anisotropic owing to the open channels. For instance, when comparing densities along different orientations of the crystal, Si$_{24}$ shows denser atomic arrangement along the $x$-axis unlike the others, which significantly differs from isotropic atomic arrangement for dSi.

We investigate thermoelectric properties of Si$_{24}$ for various temperatures up to 700 K since thermal stability of Si$_{24}$ in air was experimentally confirmed up to 750 K~\cite{Kim:2014el}. 
The ZT obtained from our calculations indicates that Si$_{24}$ shows superior thermoelectric properties compared to dSi, especially when lightly doped with electrons as shown in Figure~\ref{fig1}(c). The overall ZT values are enhanced with an increasing temperature for both Si$_{24}$ and dSi. On the other hand, enhancement factor, defined as a ratio of the maximum ZT (ZT$_{\mathrm{max}}$) between Si$_{24}$ and dSi, decreases as temperature increases, which reaches $\sim$17 at 300 K. The optimal doping concentration for the ZT$_{\mathrm{max}}$ increases with temperature as in Figure~\ref{fig1}(c). We note that overall values of the optimal doping concentration are much smaller for Si$_{24}$ than dSi. This is advantageous because impurity scattering becomes significant for heavily doped semiconductors, of which the effects are difficult to be captured directly in the first-principles calculations. 

It is worth noting here a constant relaxation time approximation can lead to inaccurate ZT value~\cite{Ouyang:2017bs}. As is known for other materials~\cite{PhysRevB.91.235419}, elaborated description of electron-phonon coupling tends to correct the overestimated ZT values obtained from simplified calculations. Likewise, the momentum dependence of relaxation time in Eq.~(\ref{eq:2}) needs to be taken into account in evaluating the electrical conductivity $\sigma$ to obtain a reliable ZT, specifically for a newly synthesized material. Another point to note is that evaluation of accurate $\sigma$ values requires remarkably dense $\boldsymbol{k}/\boldsymbol{q}$ grid points, and convergence test should be performed with caution. We confirmed that the data grid used here (40$\times$40$\times$9=14,400) gives reasonably converged ZT values less than 5\% relative error compared to that from the most dense grid mesh we tested (50$\times$50$\times$9=22,500). The detailed results are provided in SI S2.

\begin{figure}[]
	\centering
	\includegraphics[width=\columnwidth]{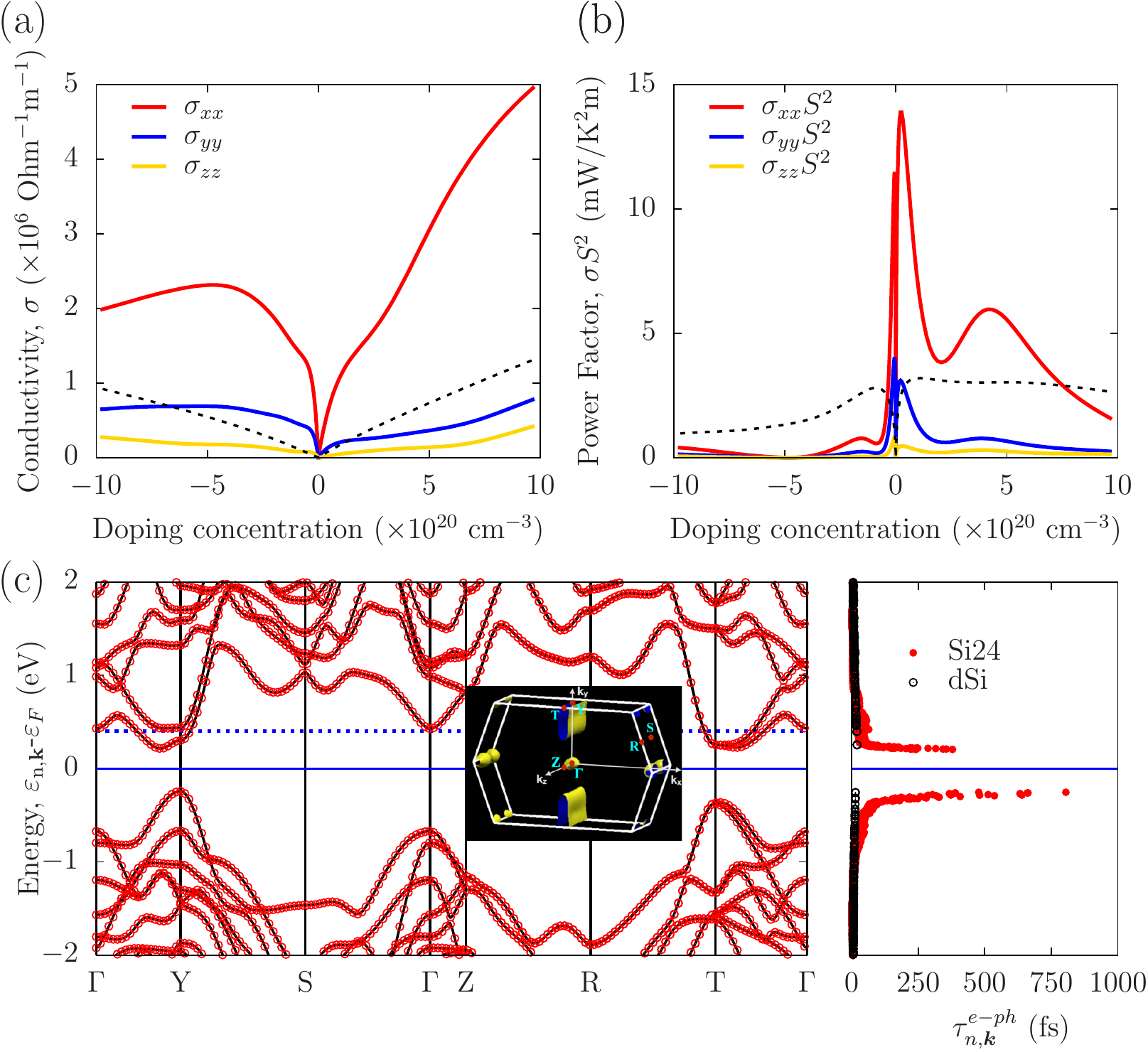}\\
	\caption{(a) Electrical conductivity tensor components and (b) power factor at 300 K with a varying doping concentration. The continuous and dashed lines are for Si$_{24}$ and dSi, respectively. (c) Electronic band dispersion of Si$_{24}$ from first-principles calculations (black lines) and maximally locallized Wannier functions (MLWFs) (red circles). Relaxation times ($\tau^{e-ph}_{n,\boldsymbol{k}}$) as in Eq.~(\ref{eq:2}) of electrons at 300 K for both Si$_{24}$ and dSi are also provided aside. In the inset, the Fermi surface in the Brillouin zone of Si$_{24}$ at a doping concentration of 2.5$\times$10$^{20}$ cm$^{-3}$, which is marked as blue dotted line in (c). In (c), the zero energy denoted by a solid horizontal line was set to be middle between conduction and valence band edges.}
	\label{fig2}
\end{figure}

To explain the enhanced ZT of Si$_{24}$, we will discuss the effects of each of the components in ZT hereafter. Firstly, we consider the effects of electronic contributions to the thermoelectric properties. We find that the electrical conductivity ($\sigma$) of Si$_{24}$ is highly anisotropic (Figure~\ref{fig2}(a)), which can be deduced from its anisotropic geometry (Figure~\ref{fig1}(a)). Compared with the dSi, it is remarkable that the $\sigma_{xx}$ for Si$_{24}$ is always higher throughout the doping range considered in this study, while $\sigma_{yy}$ and $\sigma_{zz}$ are largely suppressed. Note that the nanopores run along the $x$-axis (Figure~\ref{fig1}(b)) where the atomic arrangement is most densely packed. In contrast, along the orientations perpendicular to the nanopore axis, electrons are more likely to be scattered by electron-phonon coupling due to the anisotropic bonding. 

Similarly, power factors ($\sigma S^{2}$) also show the high anisotropy as observed in the electric conductivity (Figure~\ref{fig2}(b)). In this case, however, the magnitude of the $\sigma S^{2}$ also varies significantly due to carrier type and density. When Si$_{24}$ is lightly doped by electrons, the $\sigma_{xx} S^{2}$ reaches the maximum, significantly exceeding that of the dSi (Figure~\ref{fig2}(b)). For higher doping concentration, the $\sigma_{xx}S^{2}$ drops substantially and becomes smaller than that of dSi. For hole doping, the $\sigma S^{2}$ in all directions are significantly diminished compared to the electron-doped cases, and become much smaller than that of dSi, except the peaks at low concentration. As the power factor is linearly proportional to the ZT, this agrees well with the above observation that ZT reaches the maximum at light electron doping as in Figure~\ref{fig1}(c). 

These improved thermoelectric properties of Si$_{24}$ can be ascribed to its electronic band dispersions (Figure~\ref{fig2}(c)). With small electron doping concentration as marked in Figure~\ref{fig2}(c), there exist multiple electron pockets (or valleys) in the BZ as can be also seen in the Fermi surface at that doping level; see the inset of Figure~\ref{fig2}(c). This multiple \emph{valley degeneracy} is also known to be responsible for the good thermoelectric performance of PbTe$_{1-x}$Se$_{x}$~\cite{Pei:2011kx}. The highly anisotropic electronic structures are clearly seen in Si$_{24}$ from the Fermi surface, which is consistent with the anisotropic crystal structure. Note that the relaxation times in Si$_{24}$ are much longer than those in dSi, especially for low energies, or low doping levels. This agrees with the fact that high ZT$_{\mathrm{max}}$ of Si$_{24}$ occurs at much lower doping concentration than that of dSi as shown in Figure~\ref{fig1}(c).

\begin{figure}[]
	\centering
	\includegraphics[width=\columnwidth]{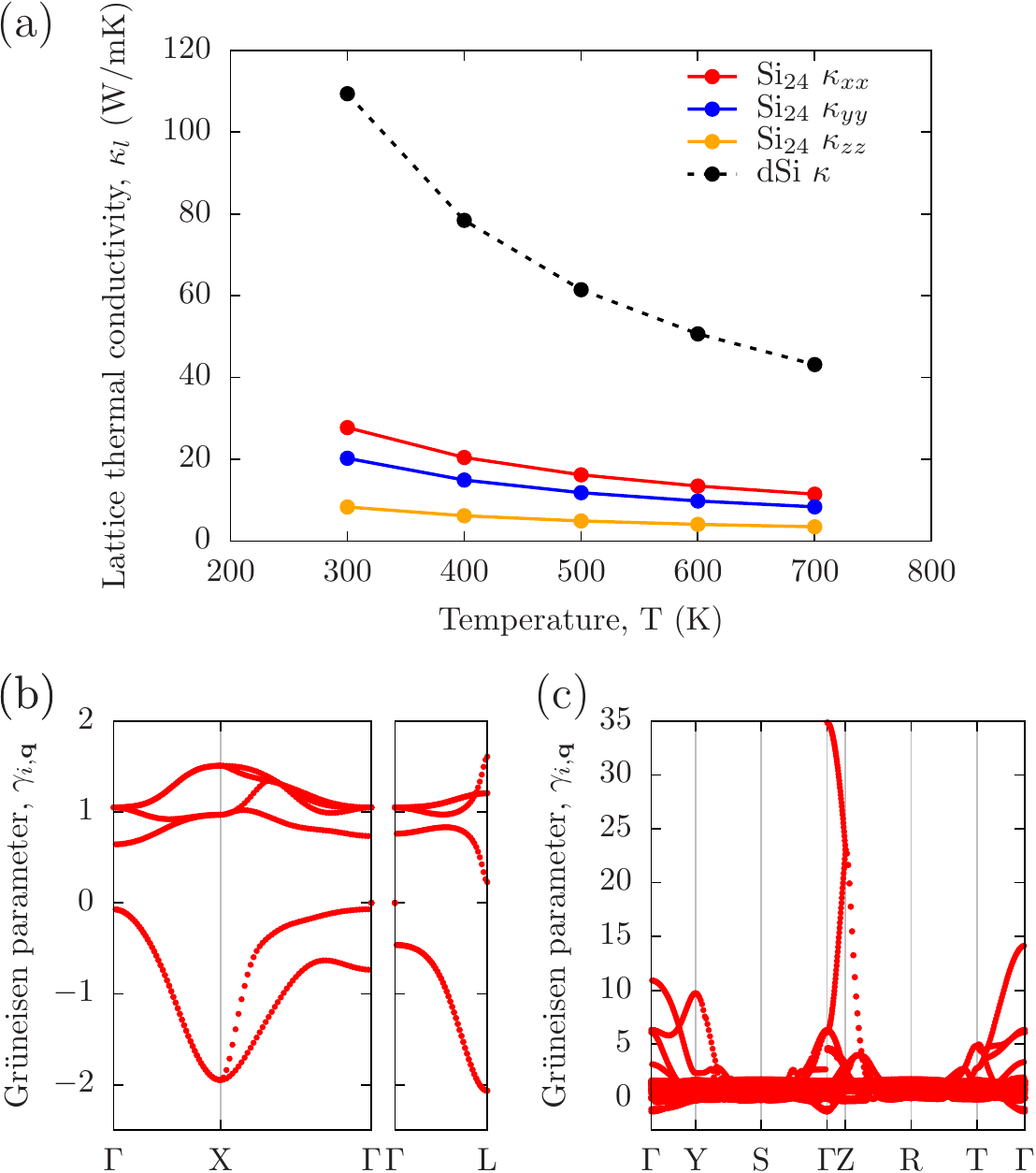}\\
	\caption{(a) Lattice thermal conductivity ($\kappa_l$) due to anharmonic phonons. Gr\"{u}neisen parameters for (b) dSi and (c) Si$_{24}$ are shown.}
	\label{fig3}
\end{figure}

Furthermore, Si$_{24}$ provides low lattice thermal conductivity ($\kappa_{l}$) as seen in Figure~\ref{fig3}(a), which renders even better thermoelectric performance compared to dSi. The calculated $\kappa_{l}$ for Si$_{24}$ is approximately 4 times lower than that of dSi along $x$-axis throughout the temperature range in this study. The difference becomes even greater to be $\sim$13 times for the $\kappa_{l}$ along $z$-axis. If we consider electronic contribution to the thermal conductivity \emph{via} Wiedemann-Franz law, the anisotropy in heat conduction becomes even greater due to higher electrical conductivity along the $x$-axis compared to the $z$-axis.

We show the Gr\"{u}neisen parameter for both crystals in Figure~\ref{fig3}(b) and (c), which is a measure of anharmonicity of the bonds. The resistivity of the heat conduction through the lattice vibrations due to the Umklapp process increases with the magnitudes of Gr\"{u}neisen parameters. The highest Gr\"{u}neisen parameters shown along the $\Gamma$-Z path indicate that the anharmonic scattering of phonons occurs mostly along the $z$-direction, and agrees with the lowest $\kappa_{l}$ along $z$-direction (Figure~\ref{fig3}(a)).

For practical aspects of its thermoelectric application, we discuss the effects of conventional defects on transport properties of Si$_{24}$.
First of all, recent experiment~\cite{PhysRevB.95.094306} and theoretical calculations~\cite{Kim:2014el} of Raman spectra agree well with each other in a wide range of temperature and pressure. From these, we could infer that the local atomic structures of synthesized Si$_{24}$ samples are quite close to the ideal ones. Moreover, a recent theoretical work on various defects in Si$_{24}$~\cite{doi:10.1021/acs.jpcc.7b04032} revealed that population of substitutional dopants are dominant over intrinsic point defects such as vacancy and interstitial thanks to their relatively lower formation energies ($<$ 0.7 eV) compared with the intrinsic ones (2.3--3.7 eV). It is noteworthy that the typical n-type dopant states with very low ionization energy hybridize with conduction bands~\cite{doi:10.1021/acs.jpcc.7b04032}, realizing good electric conductivity of the doped Si$_{24}$.

We further discuss the effects of remaining guest atoms in the cages of Si$_{24}$. As we mentioned above, the Si$_{24}$ crystal is synthesized by removing Na atoms confined in the cages of Na$_{4}$Si$_{24}$, which takes $\sim$8 days at 400 K~\cite{Kim:2014el}. This indicates that, on one hand, the Na atoms are loosely bound to the host Si atoms similar to clathrates, so that the trapped guest atoms can escape from the crystal. On the other hand, the relatively slow degassing process even at an elevated temperature is attributed to the high kinetic energy barrier for migration of Na atoms in the channel from one cage to the adjacent one. Based on our nudged elastic band calculations, the barrier is estimated to be $\sim$1 eV, which is in a good agreement to the literature~\cite{Arrieta:2017dp}. In addition, when comparing the size of the channel windows and Na atom, the transport behavior of the guest atoms is dominated by a single-file diffusion, of which the degassing rate is limited by diffusive motion of the atoms at the ends of the line. Thus, complete removal of the Na atoms might be difficult, and residual Na atoms might be remaining in the cages. 

\begin{figure}[]
	\centering
	\includegraphics[width=\columnwidth]{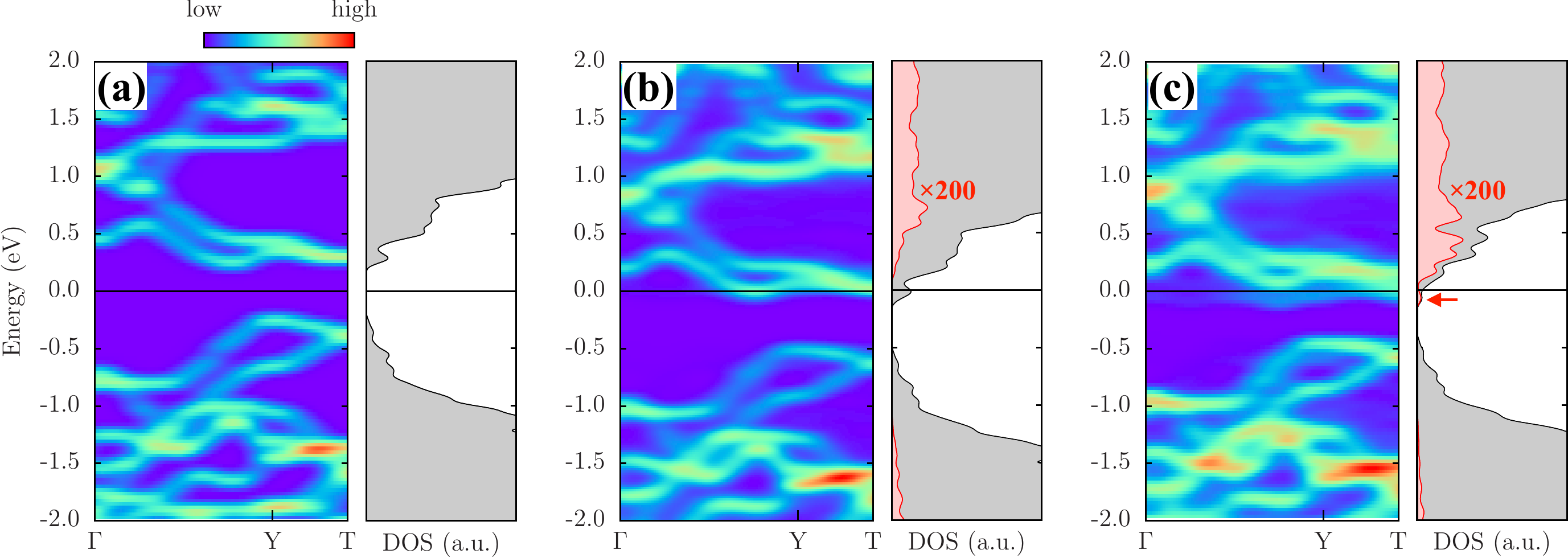}\\
	\caption{Electronic band dispersions and density of states of Na-doped Si$_{24}$ with (a) 0, (b) 1 and (c) 2 Na ions in a (3$\times$2$\times$2) supercell containing 288 Si atoms. The supercell band structures are unfolded to the irreducible Brillouin zone of the primitive cell of Si$_{24}$ by using BandUP code~\cite{PhysRevB.89.041407, PhysRevB.91.041116}. Color-coding indicates spectral weight. Total and projected DOS to Na (magnified by a factor of 200) are shown in gray and red shades, respectively. For a single Na ion content, the ionic doping concentration corresponds to 1.58$\times$10$^{20}$ cm$^{-3}$. In (a), the zero energy of the pristine Si$_{24}$ was set to be the middle between edges of conduction and valence bands and was set to be charge neutral points due to Na atom doping for both (b) and (c).}
	\label{fig4}
\end{figure}

We expect that those residual guest atoms would have the rattling effects as observed in skutterudites and clathrates~\cite{Dolyniuk:2016ht}, enhancing the thermoelectric properties of the crystal. With the guest atoms, their thermal vibrations can interact with the acoustic phonons of the host atoms, which suppresses the lattice thermal conductivity further. Similarly, thermal conductivities of carbon nanotubes (CNTs) filled with water was suppressed by 20--35~\% compared to that of empty CNT~\cite{Thomas:2010jq}. In addition, the residual Na atoms in the channel would donate electrons to the host Si$_{24}$ crystal without affecting the electronic structures. For instance, band structures in Figure~\ref{fig4} are nearly unchanged for a single Na atom out of 288 Si atoms ($\sim$0.35 at.\%), and donor level begins to appear when two Na atoms are doped ($\sim$0.7 at.\%) as indicated by the arrow. This spontaneous electron doping is advantageous for the Si$_{24}$ to be used as a TE material as discussed above (see Figure~\ref{fig1}(c and d)). Moreover, this spontaneous doping effect would significantly lower the chances of electrons being scattered by impurities (i.e., dopant atoms), which becomes dominant at heavy doping. It is worth to note that large number of guest atoms would donate many electrons to the cage structure and eventually turn the the host materials into a metal~\cite{Stefanoski:2012jf}. In this case, thermal transport is dominated by electrons, i.e., $\kappa\approx\kappa_{e}$.

\section{Conclusions}
In conclusion, we demonstrate thermoelectric properties of a new silicon allotrope, Si$_{24}$ by using various \emph{ab initio} computational methods without empirical parameters: phonon-mediated electrical conductivity and lattice thermal conductivity with anharmonic phonon effects. The highly anisotropic structure of the Si$_{24}$ results in anisotropic electronic and thermal transport properties as well. The electron-doped Si$_{24}$ displays superior thermoelectric behavior to the Si in a cubic diamond phase, which is ascribed to the enhanced power factor and reduced lattice thermal conductivity. We also pointed out that the thermoelectric performance can be further enhanced by guest atoms in the cage, due to their role as electron donator and rattler. 

\begin{acknowledgements}
We thank Korea Institute for Advanced Study for providing computing resources (KIAS Center for Advanced Computation Linux Cluster System) for this work. Y.-W.S. was supported by NRF of Korea (Grant No. 2017R1A5A1014862, SRC program: vdWMRC Center). DYK acknowledges the support by NSFC of China (Grant No. 11774015) and NRF (NRF-2017R1D1A1B03031913).
\end{acknowledgements}


%

\widetext
\clearpage

\begin{center}
\textbf{\large Supplementary Material for: \\Enhanced Thermoelectric Properties in a New Silicon Crystal Si$_{24}$ with Intrinsic Nanoscale Porous Structure} \\
\vspace{10pt}
Kisung Chae,$^{1, \ast}$ Seoung-Hun Kang,$^{1, \ast}$ Seon-Myeong Choi,$^{1, \dagger}$ Duck Young Kim,$^{2, 3, \ddagger}$ and Young-Woo Son$^{1, \S}$ \\
\vspace{4pt}
$^1$ \emph{Korea Institute for Advanced Study, Seoul 02455, South Korea} \\
$^2$ \emph{Center for High Pressure Science and Technology Advanced Research, Shanghai 201203, P. R. China} \\
$^3$ \emph{Division of Advanced Nuclear Engineering, POSTECH, Pohang, South Korea} \\
(Dated: \today)
\end{center}

\renewcommand{\thesection}{S\Roman{section}}
\setcounter{section}{0}
\renewcommand{\thefigure}{S\arabic{figure}}
\setcounter{figure}{0}
\renewcommand{\thetable}{S\arabic{table}}
\setcounter{table}{0}
\renewcommand{\bibnumfmt}[1]{[S#1]}
\renewcommand{\citenumfont}[1]{S#1}

\section{Factors on electric conductivity}
Here, we justify our assumption for electronic relaxation time, $\tau^{e}_{n,\boldsymbol{k}}\approx\tau^{e-ph}_{n,\boldsymbol{k}}$, used in our study.
According to the Matthiessen rule, scattering rates are:
$$(\tau^{e}_{n,\boldsymbol{k}})^{-1} = (\tau^{e-ph}_{n,\boldsymbol{k}})^{-1} + (\tau^{e-e}_{n,\boldsymbol{k}})^{-1} + (\tau^{imp}_{n,\boldsymbol{k}})^{-1},$$
where each of the terms on the right-hand side indicates contributions from electron-phonon, electron-electron and impurity scattering, respectively.
First of all, an elaborated \emph{ab initio} calculations~\cite{PhysRevLett.112.257402} have shown that for relaxation of excess charge carriers in silicon with cubic diamond phase generated by photon injection, electron-electron scattering is negligible, i.e., $(\tau^{e-e}_{n,\boldsymbol{k}})^{-1}\approx0$. 
Instead, it is electron-phonon interaction that plays a dominant role.
Moreover, impurity scattering also seems to have minute effects on electron transport since relatively small optimal doping concentration (e.g., 0.21$\times$10$^{20}$ cm$^{-3}$ at 300 K as in Figure 1(d)) may be readily achieved due to the spontaneous electron transfer from the guest Na ions without affecting electronic structures of the crystal as seen in Figure 4.
Furthermore, a recent theoretical study~\cite{doi:10.1021/acs.jpcc.7b04032} shows that substitutional doping generates delocalized defect states along the channel ($x$-axis), indicating that a trace of impurity, if any, would not significantly change the electronic transport behaviors: $(\tau^{imp}_{n,\boldsymbol{k}})^{-1}\approx0$.

\section{Convergence tests}

\begin{figure}[]
	\centering
	\includegraphics[width=0.6\columnwidth]{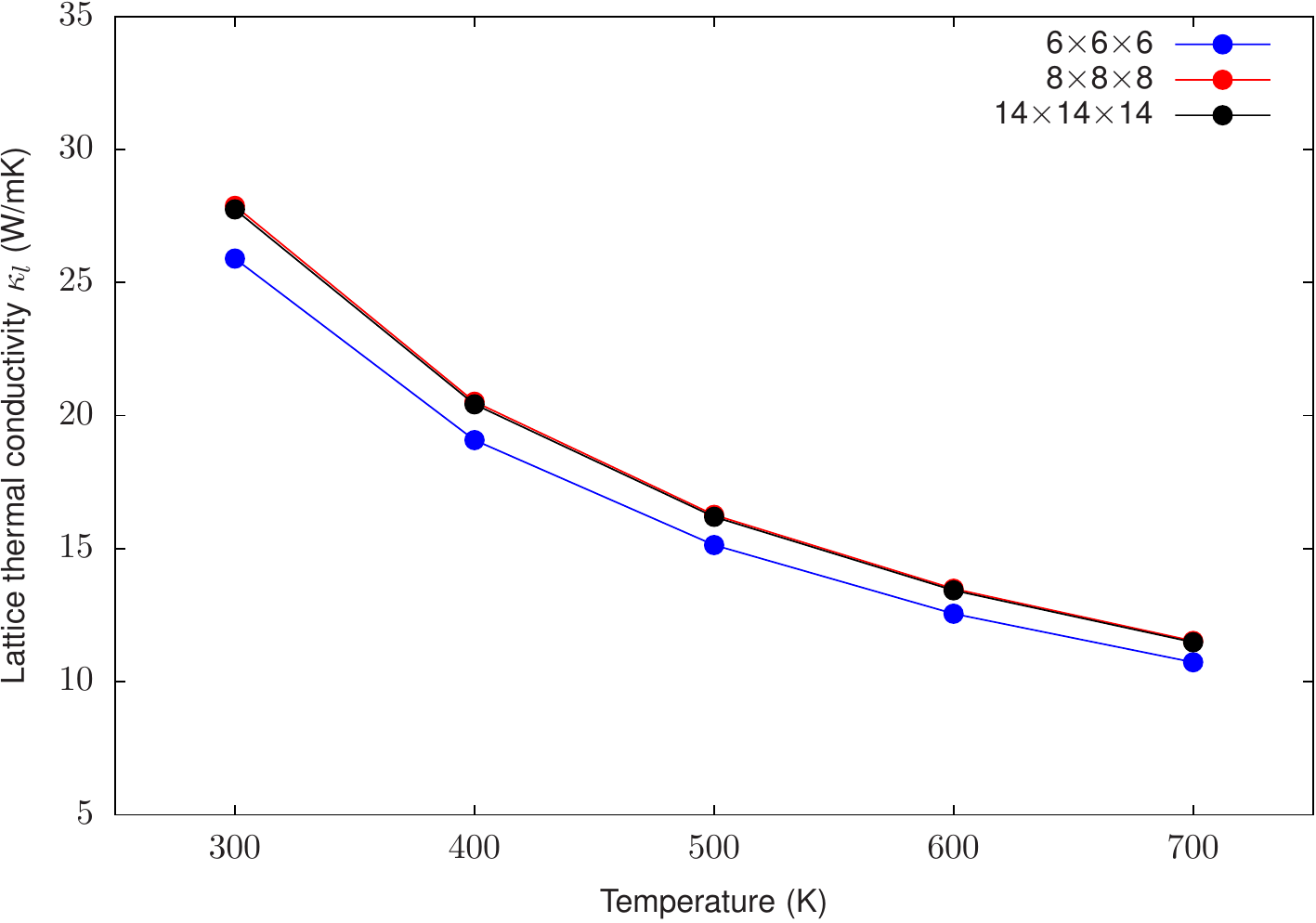}\\
	\caption{Lattice thermal conductivities ($\kappa_{xx}$) computed on different grid meshes.}
	\label{figs1}
\end{figure}

Here, we demonstrate that the computed transport coefficients in this study are obtained in well-converged conditions. 
Calculation of transport coefficients requires numerical integration on discrete grid points in the Brillouin zone, of which the accuracy sensitively depends on the grid point spacing. 
For example, for lattice thermal conductivity, coupling strength matrix of phonons needs to be dealt explicitly in the Brillouin zone~\cite{Togo:2015dy}.
Figure~\ref{figs1} shows the lattice thermal conductivity along the $x$-axis computed on different number of grid points.
The data shows that the lattice thermal conductivities seem to be converged on a 8$\times$8$\times$8 grid at all temperature range considered in this study. 
We reported the values computed on the 14$\times$14$\times$14 grid.

\begin{figure}[]
	\centering
	\includegraphics[width=\columnwidth]{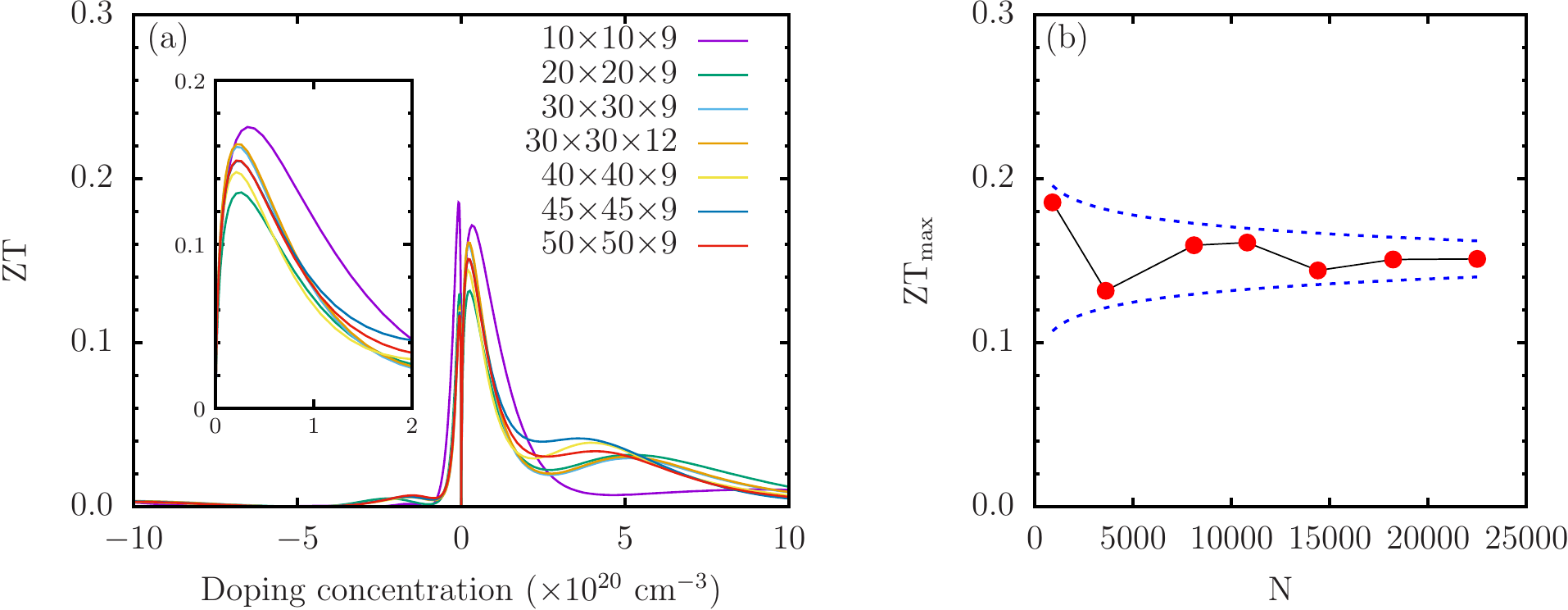}\\
	\caption{(a) Thermoelectric figure of merit (ZT) for varying doping concentration. The inset shows the ZT values due to electron doping in the vicinity of the maximum ZT values (ZT$_{\mathrm{max}}$). (b) ZT$_{\mathrm{max}}$ is plotted as a function of the number of grid points in the Brillouin zone (N). The blue dashed curves are inserted to guide the eyes.}
	\label{figs2}
\end{figure}

We also show the convergence test results on computing electron-phonon coupling with respect to the number of $\boldsymbol{k}/\boldsymbol{q}$ points in the Brillouin zone (N) in Figure~\ref{figs2}.
Note for given $\kappa$, it is convenient to compare ZT values to discuss the convergence of evaluating electron-phonon coupling as the power factor is directly proportional to ZT (eq. 2).
Firstly, it seems crucial to compute ZT with caution as the size of N affects the ZT values significantly.
For instance, on a coarse grid of (10$\times$10$\times$9), the maximum ZT occurs in hole-doped system, which is in stark contrast to all the other results computed on denser grids as in Figure~\ref{figs2}(a).
In addition, even with a demanding calculations up to the grid of (50$\times$50$\times$9), the ZT appears to converge relatively slowly with respect to N, and behaves as damped oscillation as seen in Figure~\ref{figs2}(b).
Unlike the grid spacing along the $x$- and $y$-axes, on the other hand, the grid spacing along the $z$-axis would not have significant effects on ZT when comparing the ZTs on (30$\times$30$\times$9) and (30$\times$30$\times$12).
This different behavior is attributed to the highly anisotropic electronic structures.
With a grid of (40$\times$40$\times$9) as we used in our Letter, we confirmed that the ZT$_{\mathrm{max}}$ is well-converged with the relative error less than 5\% when compared with that of (50$\times$50$\times$9).

%

\end{document}